\documentclass[12pt,titlepage,a4paper]{article}
\usepackage[cp1251]{inputenc}
\usepackage[russian,english]{babel}
\usepackage{amsmath}
\usepackage{amssymb}
\usepackage{wrapfig}
\usepackage[dvips]{graphicx}
\allowdisplaybreaks[1]

\begin{document}
\begin{center}
\Large Semileptonic decays of charmed and beauty baryons with
sterile neutrinos in the final state
\end{center}
\begin{center}
Sabir Ramazanov,

\emph {Moscow State University, 119991, Moscow, Russia\\Institute for Nuclear Research of the Russian Academy of Sciences,\\
60th October Anniversary prospect 7a, Moscow,
117312, Russia}

\end{center}
\begin{center}
\Large Abstract
\end{center}

We obtain tree-level estimates of various differential branching ratios
of heavy baryon decays with massive sterile neutrinos $\nu_{x}$ in the final state. Heavy sterile neutrinos can be searched for in heavy baryon decays 
with missing mass as a signature as well as in fixed target experiments, where heavy baryon decays contribute to sterile neutrino 
production, with sterile neutrino decays as a signature. Generally, charmed baryons are found to be less promising than charmed mesons, 
in contrast to b-hadrons. In the latter case, 
branching ratios of beauty mesons and baryons into sterile neutrinos are of the same order. As a consequence, at high energies beauty baryons give contribution 
to sterile neutrino production comparable to the contribution of beauty mesons (up to about $15\%$). 
Experimental limits on active-to-sterile mixing are quite strong for neutrinos 
lighter than D-meson 
but for heavier neutrinos they are weaker. As an example, for neutrino masses in the range $2$ GeV $\lesssim m_{\nu_{x}}\lesssim 2.5$ GeV, current data imply that the bounds on 
$\Lambda_{b}$-hyperon branching ratios into 
sterile neutrinos are ${\rm Br}(\Lambda_b\rightarrow\Lambda_c+e^-+\nu_x)\lesssim 1.3\cdot10^{-5}-1.7\cdot10^{-6}$ and 
${\rm Br}(\Lambda_b\rightarrow\Lambda_c+\mu^-+\nu_x)\lesssim 3.9\cdot10^{-7}-1.4\cdot10^{-7}$.

\section{Introduction}
Sterile neutrinos are introduced in particle physics to explain neutrino oscillations: they provide the active neutrino sector with masses and mixing. 
The inferred mass and mixing patterns in the active neutrino sector are severely constrained by neutrino experiments whereas resulting limits on parameters of the 
sterile neutrino sector 
are strongly model-dependent. In particular, sterile neutrino masses can be in the GeV range and this fact does not necessarily imply very small values of mixing 
angles \cite{Gouvea}. In 
this case sterile neutrinos can be searched for in particle physics experiments, and here we discuss in detail sterile neutrino production in baryon decays. 

For this study we assume that some sterile neutrinos have masses in the GeV range, 
and consider active-to-sterile mixing angles as free parameters constrained from direct searches for sterile neutrinos. This model-independent approach yields most 
conservative constraints
on the sterile neutrino production rates. While chosen {\it ad hoc}, the GeV range for sterile neutrino masses can be preferred in some models, with $\nu$MSM \cite{AsakaB, Asaka} serving as an example. 

The richest source of sterile neutrinos of masses in the range we consider are weak decays of heavy mesons created in beam-beam and beam-target collisions. Recently, 
these decays were revised in Ref.~\cite{Gorbunov}. If sufficiently light, sterile neutrinos are produced mostly in $\pi$- and $K$-meson decays. For heavier neutrinos, 
leptonic and semileptonic decays 
of charmed mesons are the most relevant sources of sterile neutrino production. Finally, in models with neutrinos heavier than D-mesons but still in the GeV range, decays of B-mesons dominate. 
In all cases additional contribution comes from decays of baryons. For relatively heavy neutrinos the largest contribution from this baryon channel is due 
to semileptonic decays of 
charmed and beauty baryons, among which the most promising are $\Lambda_{c}$- and $\Lambda_{b}$-hyperons because of the largest statistics collected.

In this paper we present explicit formulas for differential branching ratios of baryon semileptonic decays with massive sterile neutrinos in the final state. 
These formulas 
can be used to estimate the contribution of the baryon channel to the spectrum of sterile neutrinos produced in fix-target experiments, 
where sterile neutrino decays can be searched for. 
Likewise, the obtained formulas are relevant for searches of heavy baryon decays into sterile neutrinos which can be performed at operating and future B-factories, Tevatron and LHC.

\section{Probabilities of baryon decays}

 We start with the semileptonic decay of  $\Lambda_c$-hyperon.
 $\Lambda_c$-hyperon  weakly decays into
$\Lambda$-hyperon, positron  and electron neutrino (as well as positive muon and muon neutrino). Let sterile neutrino $\nu_{x}$ mix 
with electron neutrino $\nu_{e}$ or muon neutrino $\nu_{\mu}$ and $\theta_{lx}$ ($l=e, \mu$) is the corresponding mixing angle.
Then the amplitude of the process
$\Lambda_c\rightarrow\Lambda+l^++\nu_x$ can be written as follows,
\begin{multline}\label{matrix}
  \mathcal M = \frac{G}{\sqrt 2}V_{cs}\sin\theta_{lx} \\
  \times\bar\Lambda(P)
  \left[f_1\gamma_\nu+i\frac{f_2}{M_{\Lambda_c}}\sigma_{\nu\mu}q^\mu
    +\frac{f_3}{M_{\Lambda_c}}q_\nu
  -\left(g_1\gamma_\nu+i\frac{g_2}{M_{\Lambda_c}}\sigma_{\nu\mu}q^\mu
    +\frac{g_3}{M_{\Lambda_c}}q_\nu\right)\gamma_5\right]
  \Lambda_c(P-q)\\
    \times\bar\nu_x\gamma^\nu(1-\gamma_5)l,
\end{multline}
where $P$ and $(P-q)$ are 4-momenta of
$\Lambda$- and $\Lambda_c$-hyperons, $M_{\Lambda}$ and
$M_{\Lambda_c}$ are their masses, $G$ is the Fermi constant,
$V_{cs}$ is the element of the Cabibbo--Kobayashi--Maskawa matrix. The dimensionless form
factors $f_1$, $f_2$, $f_3$, $g_1$, $g_2$, $g_3$ entering Eq.~\eqref{matrix} 
parametrize the matrix element of the relevant hadronic current
$j^{h}_{\nu}=\bar c\gamma_\nu(1-\gamma_5)s $ between real
$\Lambda$- and $\Lambda_c$-hyperons. These form factors are functions of $q^2$.

The differential decay rate is given by
\begin{equation} \label {Sabir}
d\Gamma=\frac{\bar{|\mathcal M|}^2}{2M_{\Lambda_c}}d\Phi,
\end{equation}
where
\begin{equation}
\nonumber
d\Phi=(2\pi)^4\delta(q+k_{\nu_{x}}+k_l)
  \frac{d\vec P}{(2\pi)^32E_\Lambda}
  \frac{d\vec k_{\nu_{x}}}{(2\pi)^32E_{\nu_{x}}}
  \frac{d\vec k_{l}}{(2\pi)^32E_l},
\end{equation}
 $\vec P$, $\vec k_{\nu_{x}}$, $\vec k_l$~denote the 3-momenta of
$\Lambda$-hyperon, sterile neutrino and charged
 lepton, respectively, $E_{\Lambda}$, $E_{\nu_{x}}$, $E_{l}$~are their energies, $\bar{|\mathcal M|}^2$ is the squared amplitude, 
averaged over spins of the initial baryon and summed over spins of final particles. 

The direct calculation gives the following expression for $\bar{|\mathcal M|}^2$:
\begin{equation}\label{jest}
\begin{split}
\bar{|\mathcal M|^2}&=8G^2 |V_{cs}|^2\sin^2\theta_{lx}\
  \Bigl[(f_1^2+g_1^2)(4Pk_{\nu_{x}}\:Pk_l-2Pk_{\nu_{x}}\:qk_l-2Pk_l\:qk_{\nu_{x}})\\
&\quad-2(f_1^2-g_1^2)M_{\Lambda}M_{\Lambda_c}k_{\nu_{x}}k_l\\
&\quad-\frac{2M_{\Lambda}}{M_{\Lambda_c}}(f_1f_2+g_1g_2)
  (k_{\nu_{x}}k_l(Pq-q^2)+qk_{\nu_{x}}(Pk_l-qk_l)+qk_l(Pk_{\nu_{x}}-qk_{\nu_{x}}))\\
&\quad-2(f_1f_2-g_1g_2)(Pk_l\:qk_{\nu_{x}}+Pk_{\nu_{x}}\:qk_l+Pq\:k_{\nu_{x}}k_l)\\
&\quad-\frac{f_2^2+g_2^2}{M_{\Lambda_c}^2}
    \bigl(4Pk_{\nu_{x}}\:Pk_l\:q^2-4Pq\bigl(Pk_l\:qk_{\nu_{x}}+Pk_{\nu{x}}\:qk_l\bigr)\\
  &\qquad-2k_{\nu_{x}}k_l\bigl(2(Pq)^2-q^2M_\Lambda^2-Pq\:q^2\bigr)
    +2qk_{\nu_{x}}\:qk_l\bigl(Pq+M_\Lambda^2\bigr)\\
  &\qquad-3M_\Lambda^2q^2\:k_{\nu_{x}}k_l+4(Pq)^2k_{\nu_{x}}k_l
    -Pq\:q^2\:k_{\nu_{x}}k_l\bigr)\\
&\quad-\frac{M_{\Lambda}}{M_{\Lambda_c}}
  (f_2^2-g_2^2)(q^2\:k_{\nu_{x}}k_l+2qk_{\nu_{x}}\:qk_l)\\
&\quad+\frac{M_{\Lambda}}{M_{\Lambda_c}}(f_1f_3+g_1g_3){M_{\Lambda_c}}M_{\Lambda}
  (2qk_{\nu_{x}}(Pk_l-qk_l)+2qk_l(Pk_{\nu_{x}}-qk_{\nu_{x}})-2k_{\nu_{x}}k_l(Pq-q^2))\\
&\quad+(f_1f_3-g_1g_3)(2Pk_l\:qk_{\nu_{x}}+2Pk_{\nu_{x}}\:qk_l\:-2Pq\:k_{\nu_{x}}k_l)\\
&\quad+2\frac{f_2f_3+g_2g_3}{M_{\Lambda_c}^2}
  q^2\bigl(Pk_l\:qk_{\nu_{x}}+Pk_{\nu_{x}}\:qk_l\bigr)\\
&\quad-\frac{f_3^2}{M_{\Lambda_c}^2}(2qk_{\nu_{x}}\:qk_l-q^2\:k_{\nu_{x}}k_l)
  \bigl(Pq-M_{\Lambda}(M_{\Lambda_c}+M_{\Lambda}))\bigr)\\
&\quad-\frac{g_3^2}{M_{\Lambda_c}^2}(2qk_{\nu_{x}}\:qk_l-q^2\:k_{\nu_{x}}k_l)
  \bigl(Pq+M_{\Lambda}(M_{\Lambda_c}-M_{\Lambda}))\bigr)\\
&\quad+4\bigl(Pk_l\:qk_{\nu_{x}}-Pk_{\nu_{x}}\:qk_l\bigr)\\
  &\qquad\times\left(g_1f_1
     +\frac{g_2f_2}{M_{\Lambda_c}^2}\bigl(2Pq-q^2\bigr)
     +f_1g_2\left(1-\frac{M_{\Lambda}}{M_{\Lambda_c}}\right)
       +f_2g_1\left(1+\frac{M_{\Lambda}}{M_{\Lambda_c}}\right)\right)\Bigr].
\end{split}
\end{equation}

Integrating Eq.~\eqref{Sabir} over 3-momenta of sterile neutrino and charged lepton and over all possible  directions of the 
outgoing baryon, one  gets for the differential decay rate

\begin{equation}\label{main}
\begin{split}
\frac{d\Gamma}{dE_{\Lambda}}&=\frac{G^2|V_{cs}|^2}{64\pi^3}\frac{\sin^2\theta_{lx}}{q^4}
   \frac{\sqrt{E_{\Lambda}^2-M_{\Lambda}^2}}{M_{\Lambda_c}}
  \sqrt{\frac{q^4-2q^2(m_{\nu_x}^2+m_l^2)+(m_l^2-m_{\nu_x}^2)^2}{q^4}}\\
&\times\Bigl[\frac 13\bigl(2q^4-q^2(m_{\nu_x}^2+m_l^2)
    -(m_l^2-m_{\nu_x}^2)^2\bigr)
  \bigl((f_1^2+g_1^2)(4M_\Lambda^2q^2-12Pq\:q^2+8(Pq)^2)\\
\end{split}
\end{equation}
\begin{equation}
\begin{split}
\nonumber
&\qquad-12(f_1^2-g_1^2)M_{\Lambda}M_{\Lambda_c}q^2
  +24\frac{M_{\Lambda}}{M_{\Lambda_c}}(f_1f_2+g_1g_2)(Pq-q^2)q^2
  -24(f_1f_2-g_1g_2)Pq\:q^2\\
&\qquad-\frac{f_2^2+g_2^2}{M^2_{\Lambda_c}}
    (12Pq\:q^4-8(Pq)^2q^2+4M_\Lambda^2q^4)
  -6\frac{M_\Lambda}{M_{\Lambda_c}}(f_2^2-g_2^2)q^4\bigr)\\
&\quad+\bigl((m_l^2-m_{\nu_x}^2)^2-q^2(m_{\nu_x}^2+m_l^2)\bigr)\bigl(
  (f_1^2+g_1^2)(4M_\Lambda^2q^2+4Pq\:q^2-8(Pq)^2)\\
&\qquad-4(f_1^2-g_1^2)q^2M_\Lambda M_{\Lambda_c}
  -8\frac{M_{\Lambda}}{M_{\Lambda_c}}(f_1f_3+g_1g_3)Pq\:q^2
  -8(f_1f_3-g_1g_3)Pq\:q^2\\
&\qquad+4\frac{f_3^2}{M_{\Lambda_c}^2}q^2(Pq-M_\Lambda
(M_{\Lambda_c}+M_\Lambda))
  +4\frac{g_3^2}{M_{\Lambda_c}^2}q^2(Pq+M_\Lambda (M_{\Lambda_c}-M_\Lambda))
  \bigr)\Bigr],
\end{split}
\end{equation}
where $Pq$ and $q^2$ are functions of $E_{\Lambda}$:
\begin{align}
\nonumber
Pq&=M_{\Lambda}^2-E_{\Lambda}M_{\Lambda_c}
\end{align}
and
\begin{align}
\nonumber
q^2&=M_{\Lambda}^2+M_{\Lambda_c}^2-2M_{\Lambda_c}E_{\Lambda}.
\end{align}

Semileptonic decays of any other baryons ($B_1\rightarrow
B_2+l+\nu$) are described by similar formulas with obvious replacements: $M_{\Lambda_c}\rightarrow M_{B_{1}}$, 
$M_\Lambda\rightarrow M_{B_{2}}$, $|V_{cs}|^2\rightarrow |V_{ij}|^2$, where $V_{ij}$ is the relevant element of the Cabibbo--Kobayashi--Maskawa matrix.

\begin{figure}[!htb]
\begin{center}
\centerline{
\includegraphics[width=0.5\textwidth]{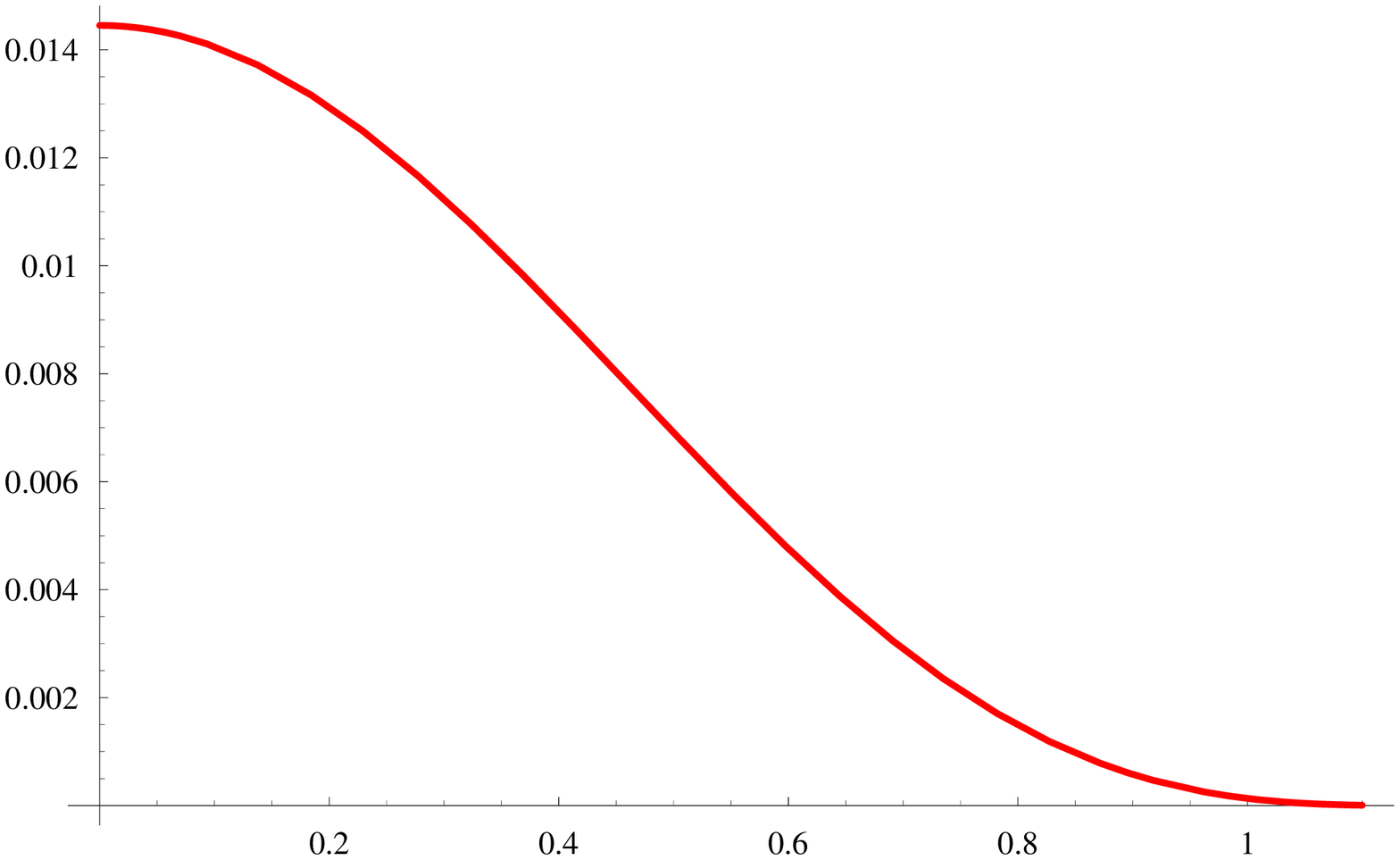}
\includegraphics[width=0.5\textwidth]{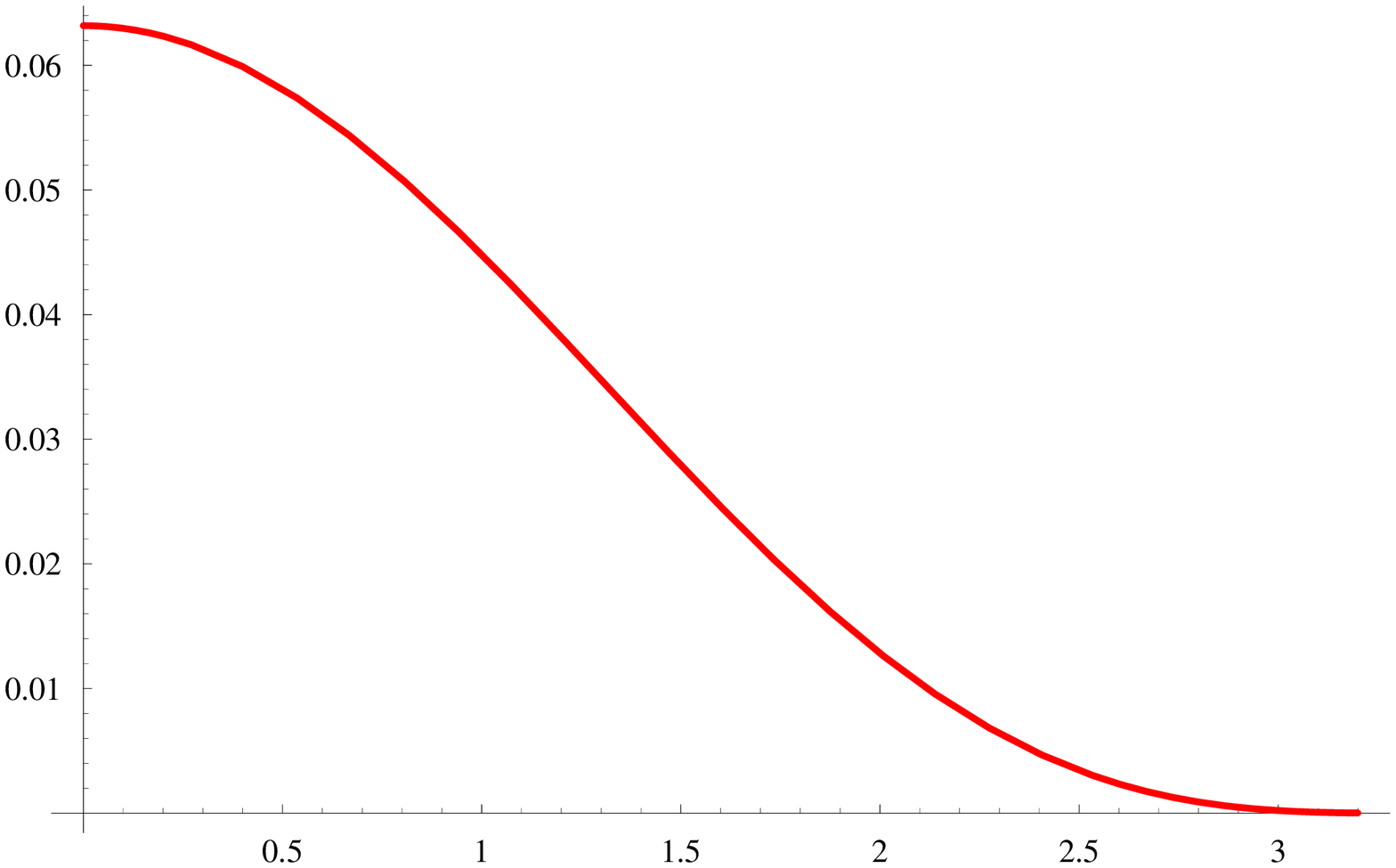}}
\begin{picture}(0,0)
\put(-135,145){a)}
\put(-135,5){$m_{\nu_{x}}$, GeV}
\put(115,5){$m_{\nu_{x}}$, GeV}
\put(100,145){b)}
\end {picture}
\caption{Branching ratios $\frac{Br}{\sin^2\theta_{lx}}$ for baryon decays: a) $\Lambda_{c}\rightarrow \Lambda +
l^++\nu_x$; b) $\Lambda_{b}\rightarrow \Lambda_{c} +
l^-+\nu_x$.\label{Br}}
\end{center}
\end{figure}

\begin{figure}[!htb]
\begin{center}
\centerline{
\includegraphics[width=0.5\textwidth]{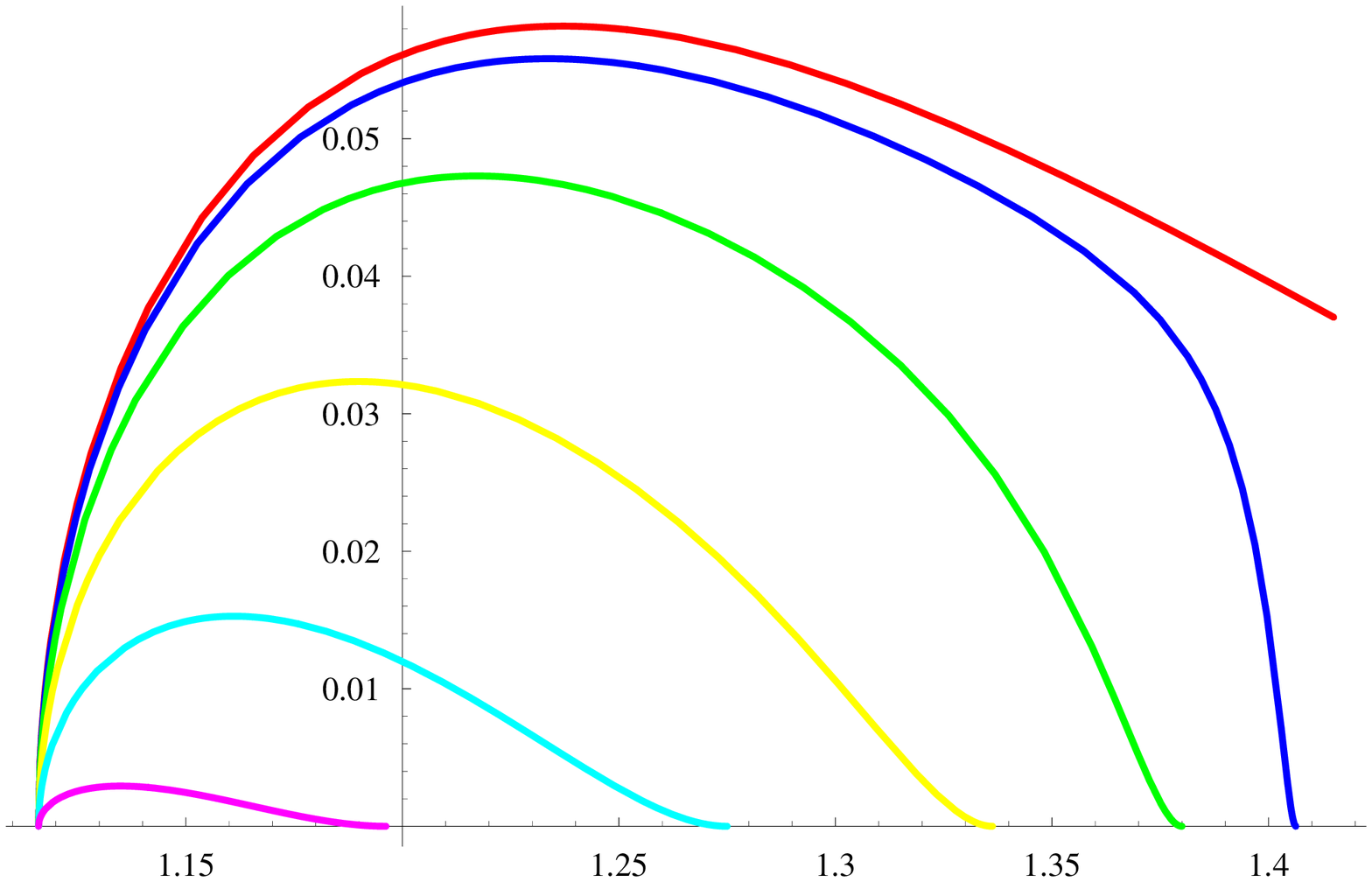}
\includegraphics[width=0.5\textwidth]{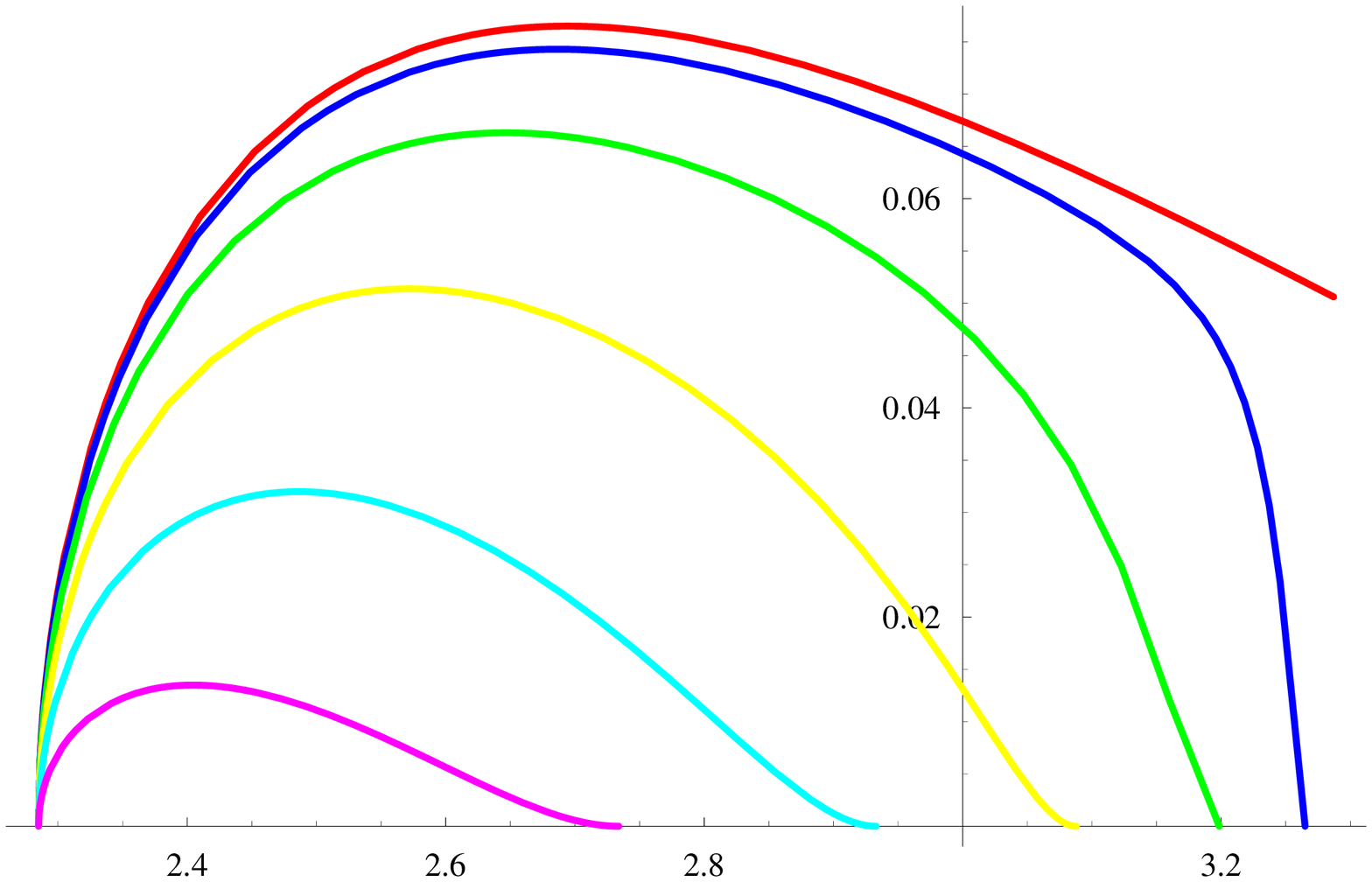}}
\begin{picture}(0,0)
\put(-135,165){a)}
\put(-135,5){$E_{\Lambda}$, GeV}
\put(115,5){$E_{\Lambda_{c}}$, GeV}
\put(100,165){b)}
\end{picture}
\caption{Differential branching ratios $\frac{d(Br/\sin^2\theta_{lx})}{d(E_{\Lambda}/{\rm GeV})}$ as functions of final baryon energy for baryon decays: a)
$\Lambda_{c}\rightarrow \Lambda + l^++\nu_x$, various plots
correspond to various sterile neutrino masses: from top to bottom $m_{\nu_x}$=0, 0.2, 0.4, 0.6, 0.8, 1.0 GeV; b)
$\Lambda_{b}\rightarrow \Lambda_{c} + l^-+\nu_x$; $m_{\nu_x}$=0, 0.5, 1.0, 1.5, 2.0, 2.5 GeV. \label{absurd}}
\end{center}
\end{figure}

By integrating Eq.~\eqref{main} over the final baryon energy one obtains the decay rate $\Gamma$ as the function of
sterile neutrino mass $m_{\nu_{x}}$. The corresponding branching ratios of the decays $\Lambda_{c}\rightarrow \Lambda + l^++\nu_x$ and
$\Lambda_{b}\rightarrow \Lambda_{c} + l^-+\nu_x$ are presented in
Fig.~\ref{Br}; charged lepton is considered to be
massless (electron or positron). Results of numerical calculations for the differential
branching ratios, which give differential spectrum of outgoing baryons, are presented in Fig.~\ref{absurd}.

In all numerical calculations we use the form
factors in the dipole approximation \cite{Cheng}:
\begin{align}
f_i(q^2)&=\frac{f_i(0)}{\left(1-\frac{q^2}{m_V^2}\right)^2},\\
g_i(q^2)&=\frac{g_i(0)}{\left(1-\frac{q^2}{m_A^2}\right)^2},
\end{align}
where for charmed baryons  $m_V=2.11$ GeV,
$m_A=2.54$ GeV, and for beauty baryons $m_V=6.34$ GeV, $m_A=6.73$
GeV. Values of $f_i(0)$ and $g_i(0)$ for the transitions $ \Lambda_{b}
\rightarrow \Lambda_{c}$ and $ \Lambda_{c} \rightarrow \Lambda $ are taken from Ref.~\cite{Cheng}. They are summarized in Table ~\ref{muchacha}. 
\begin{table}[!htb]
\hspace{5cm}
\begin{tabular}{|c|c|c|}
\hline
Form factors & $ \Lambda_{b}\rightarrow \Lambda_{c}$ & $ \Lambda_{c} \rightarrow \Lambda $\\
\hline
 $ f_{1}(0) $ & 0.53 & 0.29\\
\hline
 $ f_{2}(0) $ & 0.12 & 0.14\\
\hline
 $ f_{3}(0) $ & 0.02 & 0.03\\
\hline
 $ g_{1}(0) $ & 0.58 & 0.38\\
\hline
 $ g_{2}(0) $ & 0.02 & 0.03\\
\hline
 $ g_{3}(0) $ & 0.13 & 0.19\\
\hline
\end{tabular}
\caption{
\label{muchacha}
Form factors for the transitions $\Lambda_{c}\rightarrow\Lambda$ and $\Lambda_{b}\rightarrow\Lambda_{c}$ adopted from Ref.~\cite{Cheng} and used in numerical calculations.}
\end{table}
Values of $f_2(0)$, $f_3(0)$, $g_2(0)$, $g_3(0)$ have 
the opposite sign as compared to Ref.~\cite{Cheng} because of the different parametrization of matrix element \eqref{matrix}.

\vspace{0.5cm}

\begin{figure}[!htb]
\begin{center}
\centerline{
\includegraphics[width=0.5\textwidth]{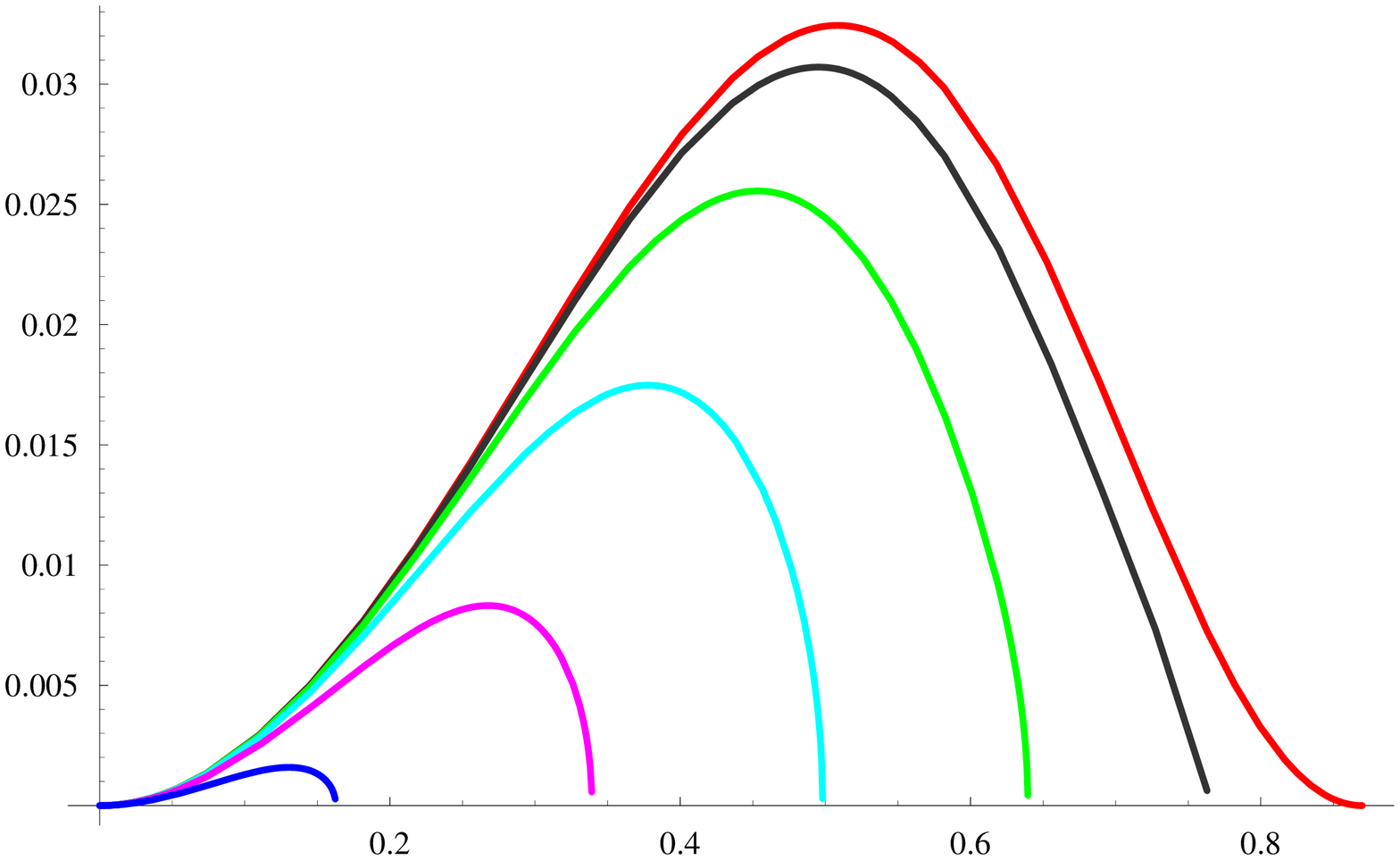}
\includegraphics[width=0.5\textwidth]{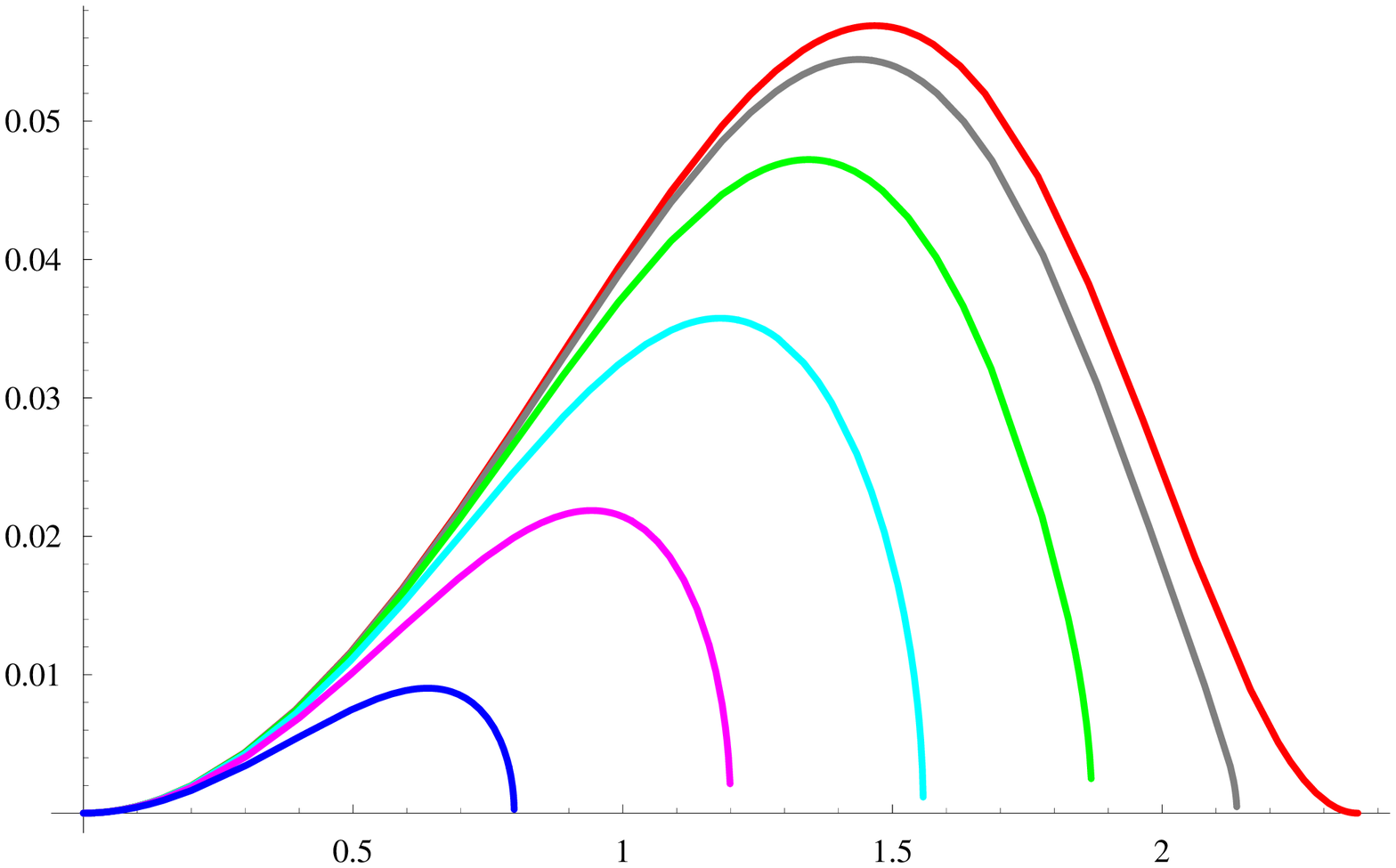}}
\begin{picture}(0,0)
\put(-160,165){a)} 
\put(-135,5){$E_{l}$, GeV}
\put(115,5){$E_{l}$, GeV}
\put(100,165){b)}
\end{picture}
\caption{Differential branching ratios $\frac{d(Br/\sin^2\theta_{lx})}{d(E_{l}/{\rm GeV})}$ as functions of charged lepton energy for baryon decays: a)
$\Lambda_{c}\rightarrow \Lambda + l^++\nu_x$;  b)
$\Lambda_{b}\rightarrow \Lambda_{c} + l^-+\nu_x$. Sterile neutrino
masses are the same as in Fig.~\ref{absurd}. Larger branching ratios correspond to smaller neutrino masses.\label{losos}}
\end{center}
\end{figure}

The differential rate $\frac{d\Gamma}{dE_{l}}$, describing the differential spectrum of outgoing charged leptons, is obtained by integrating Eq.~\eqref{Sabir} over  
$\vec P$, $\vec k_{\nu_{x}}$. The final expression is:
\begin{equation} \label{El}
\frac{d\Gamma}{dE_{l}}= \frac{1}{64\pi^3
M_{\Lambda_c}}\sqrt{E^2_l - m^2_l}\sqrt{\frac{(p^2 +
M^2_{\Lambda} - m^2_{\nu_{x}})^2}{4p^4} -
\frac{M^2_{\Lambda}}{p^2}}\int{\bar{|{\mathcal M}|^2}\sin{\theta}d\theta},
\end{equation}
where
\begin{equation}
\nonumber
p^2=M^2_{\Lambda_{c}}+m^2_{l}-2M_{\Lambda_{c}}E_{l},
\end{equation}
$\theta$ is the angle between charged 
lepton and outgoing baryon in the center-of-mass frame of sterile neutrino and outgoing baryon. The results of numerical calculation are presented in Fig.~\ref{losos}. 
Differential spectrum \eqref{El} can be used in searches of heavy baryon decays with sterile neutrinos in the final state.

\begin{figure}[!htb]
\begin{center}
\centerline{
\includegraphics[width=0.5\textwidth]{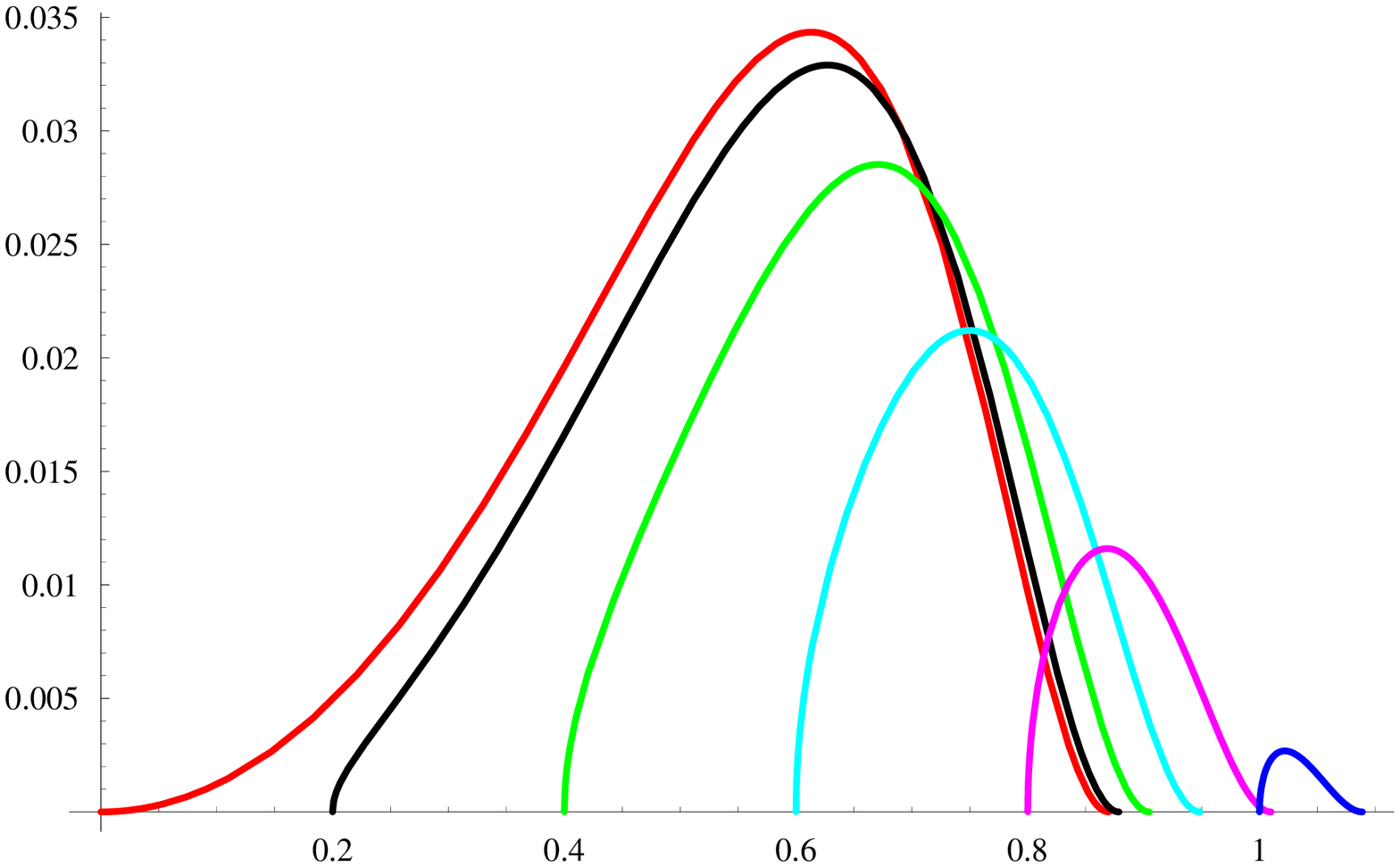}
\includegraphics[width=0.5\textwidth]{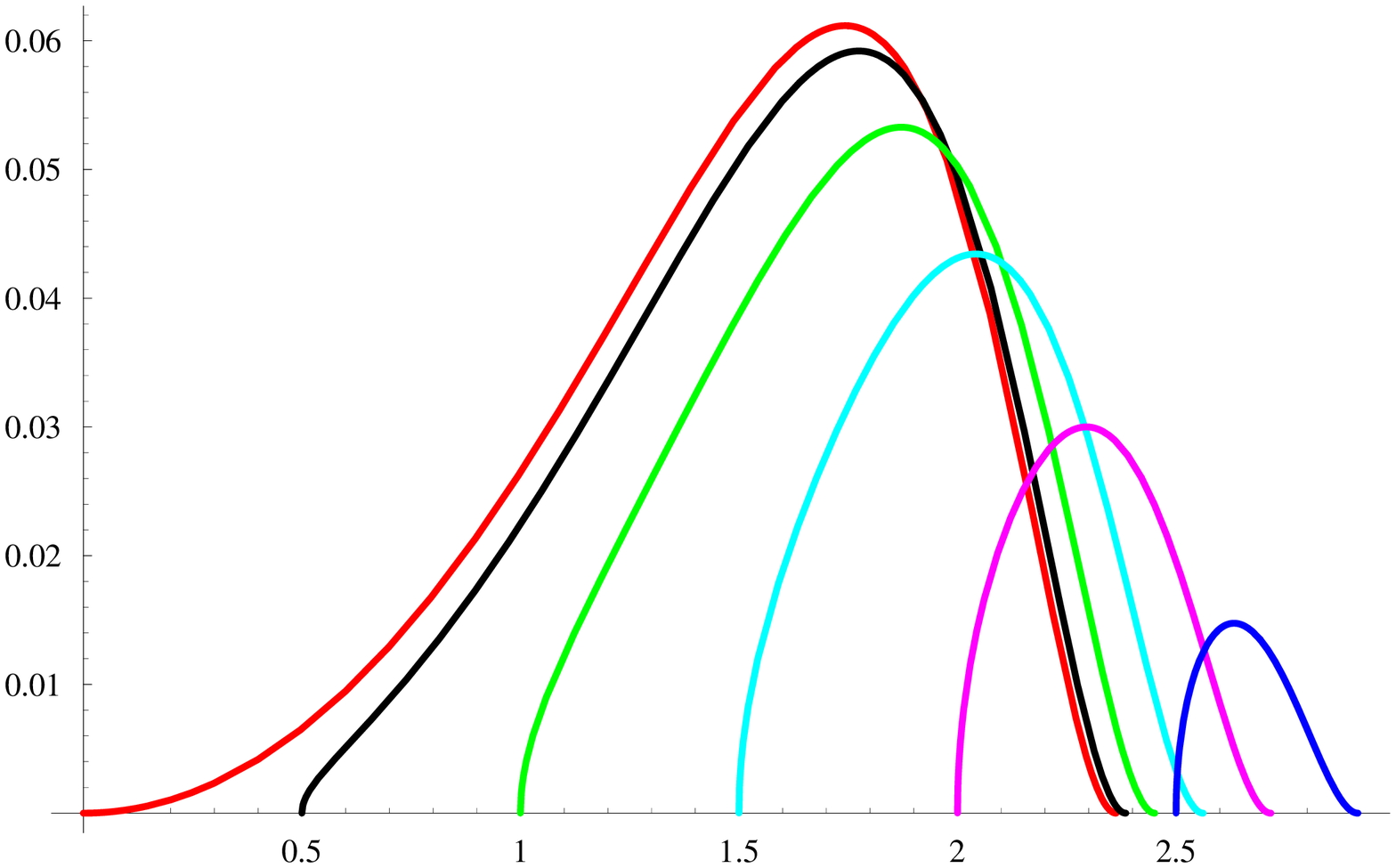}}
\begin{picture}(0,0)
\put(-135,165){a)}
\put(-135,5){$E_{\nu_x}$, GeV}
\put(115,5){$E_{\nu_x}$, GeV}
\put(100,165){b)}
\end{picture}
\caption{Differential branching ratios $\frac{d(Br/\sin^2\theta_{lx})}{d(E_{\nu_{x}}/{\rm GeV})}$ as functions of sterile neutrino energy for baryon decays: a) $\Lambda_{c}\rightarrow \Lambda + l^++\nu_x$;  b) $\Lambda_{b}\rightarrow \Lambda_{c} + l^-+\nu_x$. 
Sterile neutrino masses are the same as in  Fig.~\ref{absurd}. Larger branching ratios correspond to smaller neutrino masses.\label{zvezdi}}
\end{center}
\end{figure}

The differential rates $\frac{d\Gamma}{dE_{\nu_{x}}}$ are given by the same expression with obvious replacement: $\nu_{x}\leftrightarrow l$; $\theta$ is now the angle
between neutrino and outgoing baryon in the center-of-mass frame of lepton and outgoing baryon. The numerical results are presented in Fig.~\ref{zvezdi} and are relevant 
for searches of sterile neutrinos in baryon decays, since $\frac{d\Gamma}{dE_{\nu_{x}}}$ describes the spectrum of produced neutrinos.

\section{Discussion}
Our results for the baryon branching ratios into sterile neutrinos can be compared to similar results 
for heavy meson decays, presented in Ref.~\cite{Gorbunov}. In the case of charmed hadrons the general conclusion is that baryon branching ratios 
are always significantly lower than branching ratios of charmed mesons. As an example, for 
$m_{K}\lesssim m_{\nu_{x}}\lesssim 1$ GeV, where $m_K$ is K-meson mass, one estimates 
$\frac{Br(D\rightarrow K^0+l+\nu_{x})}{Br(\Lambda_c\rightarrow \Lambda +l^++\nu_x)}\simeq 10$. Branching ratios of leptonic meson decays are even larger. Hereafter, 
meson branching ratios are calculated with the use of formulas presented in Ref.~\cite{Gorbunov} and form factors presented in Ref.~\cite{Stech}. Hence, 
charmed baryons are less promising for sterile neutrino searches in comparison with charmed mesons, since larger statistics is required 
to reach the same level of statistical sensitivity to active-to-sterile mixing angles. From the analyses of baryon and meson decay modes, presented in this work and in 
Ref.~\cite{Gorbunov}, respectively, 
and with account of baryon fraction in c-quark hadronization \cite{myaso} one can estimate the contribution of baryon channel to sterile neutrino production in proton 
beam-beam and beam-target collisions at the level of $2\%$.

In beauty sector the situation is quite the opposite: baryon branching ratios are somewhat larger than branching ratios of mesons. As an example, for 
$2$ GeV $\lesssim m_{\nu_{x}}\lesssim 3$ GeV we obtain 
$\frac{Br(B\rightarrow D^*+l+\nu_{x})}{Br(\Lambda_b\rightarrow \Lambda_{c}+l^-+\nu_{x})}\simeq 0.5$. Note that $B\rightarrow D^*+l+\nu_{x}$ dominates the sterile neutrino 
production in meson channel. Thus, one expects that searches for sterile neutrinos in baryon decays should be competitive with similar searches in meson decays. 
From the analyses of baryon and meson decay modes, presented in this work and in Ref.~\cite{Gorbunov}, respectively, and with account of baryon fraction in b-quark 
hadronization \cite{pdg} one can estimate the contribution of baryon channel to sterile neutrino production in proton beam-beam and beam-target collisions at the level of $15\%$.

The absolute values of baryon branching ratios into sterile neutrinos are propotional to squared values of 
corresponding active-to-sterile mixing angles. The latter are limited from above 
due to negative results of direct searches for sterile neutrinos. For mixing with electron neutrinos, the strongest limits 
are $|\theta_{ex}|^{2}\lesssim 3\cdot10^{-6}-3\cdot10^{-7}$ for $m_{K}\lesssim m_{\nu_{x}}\lesssim 1$ GeV from BEBC 
experiment \cite{bebc}, $|\theta_{ex}|^{2}\lesssim 10^{-7}$ for $1.5$ GeV $\lesssim m_{\nu_{x}}\lesssim 2$ GeV 
from CHARM \cite{charm} and $|\theta_{ex}|^{2}\lesssim 1\cdot10^{-3}-1\cdot10^{-4}$ for $2$ GeV $\lesssim m_{\nu_{x}}\lesssim 3$ GeV from HRS \cite{hrs}. 
For mixing with muon neutrinos the limits are $|\theta_{\mu x}|^{2}\lesssim 5\cdot10^{-7}-1\cdot10^{-7}$ for $m_{K}\lesssim m_{\nu_{x}}\lesssim 1$ GeV, 
$|\theta_{\mu x}|^{2}\lesssim 6\cdot10^{-8}-1\cdot10^{-7}$ for $1.5$ GeV $\lesssim m_{\nu_{x}}\lesssim 2$ GeV from NuTeV \cite{nutev}, 
$|\theta_{\mu x}|^{2}\lesssim 3\cdot10^{-5}-4\cdot10^{-5}$ for $2$ GeV $\lesssim m_{\nu_{x}}\lesssim 2.5$ GeV from CHARM II experiment\cite{charm ii} 
and $|\theta_{\mu x}|^{2}\lesssim 5\cdot 10^{-4}-1\cdot10^{-4}$ from HRS experiment \cite{hrs}.

Consequently, in models with sterile neutrino mass 
ranging within $m_{K}\lesssim m_{\nu_{x}}\lesssim 1$ GeV one estimates from Fig.~\ref{Br} and the above limits:
\begin{equation}
\nonumber
\begin{split}
&\quad Br(\Lambda_c\rightarrow \Lambda+e^++\nu_{x})\lesssim 2.1\cdot10^{-8},\\
&\quad Br(\Lambda_c\rightarrow \Lambda+\mu^++\nu_{x})\lesssim 3.5\cdot10^{-9}.
\end{split}
\end{equation}

In models with $1.5$ GeV $\lesssim m_{\nu_{x}}\lesssim 2$ GeV:
\begin{equation}
\nonumber
\begin{split}
&\quad Br(\Lambda_b\rightarrow \Lambda_{c}+e^-+\nu_{x})\lesssim 2.8\cdot10^{-9}-1.3\cdot10^{-9},\\
&\quad Br(\Lambda_b\rightarrow \Lambda_{c}+\mu^-+\nu_{x})\lesssim 1.7\cdot10^{-9}-1.3\cdot10^{-9},
\end{split}
\end{equation}
where larger value on the right hand side corresponds to smaller value of $m_{\nu_{x}}$ and vice versa. 

In models with $2$ GeV $\lesssim m_{\nu_{x}}\lesssim 2.5$ GeV:
\begin{equation}
\nonumber
\begin{split} 
&\quad Br(\Lambda_b\rightarrow \Lambda_{c}+e^-+\nu_{x})\lesssim 1.3\cdot10^{-5}-1.7\cdot10^{-6},\\
&\quad Br(\Lambda_b\rightarrow \Lambda_{c}+\mu^-+\nu_{x})\lesssim3.9\cdot10^{-7}-1.5\cdot 10^{-7}.
\end{split}
\end{equation}
 
Finally, in models with heavy sterile neutrinos in the range $2.5$ GeV $\lesssim m_{\nu_{x}}\lesssim 3$ GeV:
\begin{equation}
\nonumber
\begin{split}
&\quad Br(\Lambda_b\rightarrow \Lambda_{c}+e^-+\nu_{x})\lesssim 1.7\cdot10^{-6},
\end{split}
\end{equation}
and the same limit for 
the process $\Lambda_b\rightarrow \Lambda_{c}+\mu^-+\nu_{x}$.

Thus, to search for sterile neutrinos, one has to collect statistics of at least a few million heavy baryons.

\section{Acknowledgements}
The author thanks D. S. Gorbunov and V. A. Rubakov for useful discussions.


\vspace{20cm}

\begin{thebibliography}{99}
\bibitem{Gouvea}A. de Gouvea, NUHEP-TH-07-06 (2007) [arXiv:hep-ph/0706.1732].
\bibitem{AsakaB}T. Asaka, S. Blanchet and M. Shaposhnikov, Phys. Lett. B \textbf{631} (2005) 151 [arXiv:hep-ph/0503065].\label{AsakaB}
\bibitem{Asaka}T. Asaka and M. Shaposhnikov, Phys. Lett. B \textbf{620} (2005) 17 [arXiv:hep-ph/0505013].\label{Asaka}
\bibitem{Gorbunov}D. Gorbunov and M. Shaposhnikov, JHEP \textbf{0710} (2007) 015 [arXiv:hep-ph/0705.1729].\label{Gorbunov}
\bibitem{Cheng}H. Y. Cheng and B. Tseng, Phys. Rev. D \textbf{53} (1996) 1457; D \textbf{55} (1997) 1697 (E).\label{Cheng}
\bibitem{Stech}D. Melikhov and B. Stech, Phys. Rev. D \textbf{62} (2000) 014006 [arXiv:hep-ph/0001113].\label{Stech}
\bibitem{myaso}C. Lourenco and H. K. Worhy, Phys. Rept. \textbf{433} (2006) 127 [arXiv:hep-ph/0609101].\label{myaso}
\bibitem{pdg}C. Amsler et al. [Particle Data Group], Phys. Lett. B \textbf{667} (2008) 1.\label{pdg}
\bibitem{bebc}A. Cooper--Sarkar et al., Phys. Lett. B \textbf{122} (1985) 207.\label{bebc}
\bibitem{charm}J. Dorenbosch et al. [CHARM collaboration], Phys. Lett. B \textbf{166} (1986) 473.\label{charm}
\bibitem{hrs}C. Akerlof et al. [HRS Collaboration], Phys. Rev. D \textbf{37} (1986) 577.\label{hrs}
\bibitem{nutev}A. Vaitaitis et al. [NuTeV Collaboration], Phys. Rev. Lett. (1999) 4943 [arXiv:hep-ex/9708011].\label{nutev}
\bibitem{charm ii}P. Vilain et al. [CHARM II Collaboration], Phys. Lett. B \textbf{351} (1995) 357.\label{charm ii}


\end{thebibliography}
\end{document}